\documentclass[a4paper]{mem}
\usepackage{natbib}
\usepackage{graphicx}
\usepackage[a4paper]{hyperref}
\idline{???}{1}
\begin{document}
   \title{The role of the Initial Mass Function in modelling the Intra--Cluster
          Medium
}

   \author{L.\ Portinari}


   \institute{Theoretical Astrophysics Center, Juliane Maries Vej 30,
DK-2100 Copenhagen \O \\
 \email{lportina@tac.dk} 
             }

   \abstract{
The expected metal enrichment of the intra--cluster medium (ICM) and the 
partition of metals between cluster galaxies and the hot ICM depends on the 
stellar Initial Mass Function (IMF). The choice of the IMF in simulations 
of clusters has also important consequences on the ``cold fraction'', 
which is a fundamental constraint on cluster physics.

We discuss the chemical enrichment and the cold fraction in clusters as 
predicted with different IMFs, by means of a straightforward approach that is 
largely independent of the details of chemical evolution models or simulations.
We suggest this simple approach as a guideline to select the input parameters 
and interpret the results of more complex models and hydrodynamical 
simulations. 
   }
   \authorrunning{L.\ Portinari}
   \titlerunning{The role of the IMF in modelling the ICM}
   \maketitle
%

\begin{figure}
\centering
\resizebox{0.5 \textwidth}{!}{\rotatebox[]{-90}{\includegraphics{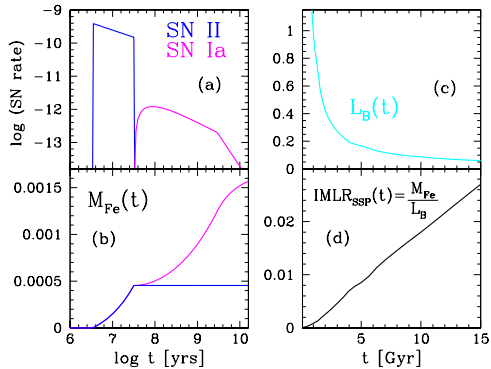}}}
\caption{{\bf(a)} Rates of SN~II and SN~Ia (number per year) for a 
1~$M_{\odot}$ Salpeter SSP.
{\bf (b)} Cumulative iron production from SN~II and SN~Ia.
{\bf (c)} Luminosity evolution of the SSP. 
{\bf (d)} Evolution of IMLR$_{SSP}$.} 
\label{fig:IMLR_SSP}
\end{figure}

\section{Introduction}
The chemical enrichment of the intra--cluster medium (ICM) has been
extensively discussed in literature by means of chemical evolution models
of elliptical galaxies with galactic winds, and a variety of scenarios
has been advanced (see the review by Matteucci, these proceedings). 
Very recently, cosmological hydro--dynamical simulations 
of cluster formation have been developed, that can follow self--consistently 
the star formation and chemical enrichment history of cluster galaxies 
and of the ICM (Valdarnini 2003; Tornatore et al.\ 2004; Romeo et~al.\ 2004).
Star formation has important effects 
on the hydro--dynamical evolution of the cluster, via energy feedback from 
supernov\ae\ and metal enrichment of the ICM: the first effect contrasts, 
the second one boosts the cool-out of the hot gas. The chemical evolution of
the ICM is not only an interesting issue {\it per se},
but an important ingredient of the global physical evolution of clusters. 
As a consequence, the choice of input parameters and ``recipes'' related 
to star formation (notably,
the stellar Initial Mass Function and the implementation of sub--grid feedback
effects) is crucial for the results of the simulations.

In this paper we outline a simple procedure to estimate 
the chemical enrichment of the ICM, the partition of the 
metals between stars and ICM, and the cold fraction expected after an assumed
Initial Mass Function (IMF). This provides
a guideline to select the optimal input parameters of the simulation; 
and to distinguish, in the results of a complex and fully self--consistent
simulation, what is merely a consequence of the adopted IMF, and what is 
an effect of the interplay between hydrodynamical evolution and
star formation. 

\section{IMF, metal production and partition between stars and ICM}
Let's choose an IMF and consider a burst of star formation (a Single Stellar 
Population, SSP) with stellar masses distributed accordingly; we can compute 
the expected rates of SN~II and SN~Ia, the corresponding rate of production 
of metals (for example, iron: $M_{Fe}(t)$), and the luminosity evolution
of the SSP $L_B(t)$. We can then define the Iron Mass--to--Light ratio typical
of that IMF as IMLR$_{SSP} (t) = \frac{M_{Fe}}{L_B}$. Fig.~\ref{fig:IMLR_SSP} 
illustrates this procedure for the Salpeter (1955) IMF with mass limits 
[0.1--100]~M$_{\odot}$; the detailed calculations can be found in 
Portinari et~al.\ (2004, hereafter PMCS).

Part of the iron produced will be ``eaten up'' by subsequent 
star formation episodes, to build up the stellar metallicities observed in 
cluster galaxies; only the remaining fraction is available to enrich the ICM. 
We need to estimate how the total iron produced gets partitioned
between the stellar populations and the ICM (Renzini et~al.\ 1993). 

\begin{figure}
\centering
\resizebox{0.35 \textwidth}{!}{\rotatebox[]{-90}{\includegraphics{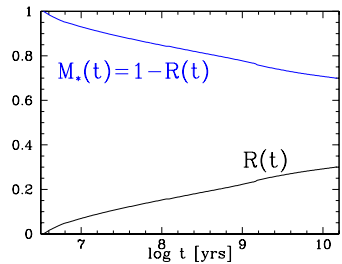}}}
\caption{Mass fraction $R$ of returned gas from dying stars and 
complementary locked--up fraction $M_*$ for a Salpeter SSP.}
\label{fig:returned}
\end{figure}

We can estimate the amount of iron locked in the stars as 
{\mbox{$M_{Fe,*}=Z_{Fe,*} \times M_*$}}. As to the stellar metallicity
$Z_{Fe,*}$, the star mass in clusters is dominated by massive 
ellipticals, with global metallicities between 0 (solar) and +0.2~dex, 
and [$\alpha$/Fe] ratios around +0.2~dex (PMCS and references therein).
The mass in stars $M_*$ is not directly observable; it is usually inferred
from the global luminosity of cluster galaxies, but the conversion factor,
the stellar Mass--to--Light ratio ($M_*/L$), depends on the IMF. Thus,
the mass $M_*$ locked in stars is best computed self--consistently 
from the assumed IMF (Fig.~\ref{fig:returned}); $M_*$ drives the actual amount
of iron $M_{Fe,*}$ required to reproduce the observed metallicity $Z_{Fe,*}$.

\begin{figure}
\centering
\resizebox{0.45 \textwidth}{!}{{\includegraphics{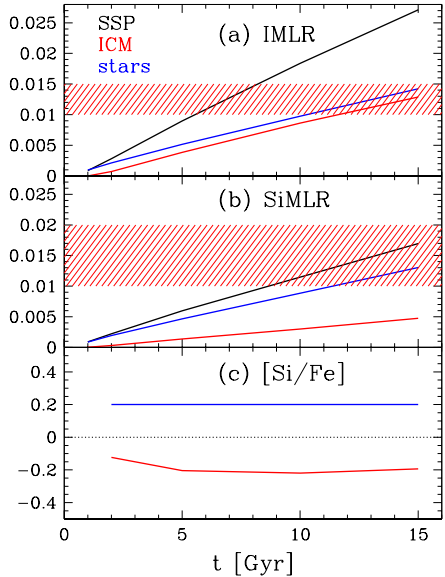}}}
\caption{{\bf (a)} IMLR$_{SSP}$ for the Salpeter IMF split into its
components: IMLR$_*$ locked in the stars and IMLR$_{ICM}$ in the ICM; 
the red shaded area is the observed IMLR in the ICM (Finoguenov et~al.\ 2000,
2001).
{\bf (b)} Same as (a), for the Silicon Mass--to--Light ratio. 
{\bf (c)} [Si/Fe] ratio in the stars (assumed {\it a priori}) and in the ICM.}
\label{fig:IMLR_SiMLR_SiFe}
\end{figure}

We are now able to split the characteristic IMLR$_{SSP}$ of the assumed IMF 
into the fraction IMLR$_* = \frac{M_{Fe,*}}{L_B}$ locked in the stellar 
populations, and the remaining fraction IMLR$_{ICM}$ that can be compared 
to ICM observations (Fig.~\ref{fig:IMLR_SiMLR_SiFe}a, red line vs.\ red shaded
area). Notice that IMLR$_{ICM}$ 
as computed here is an {\it upper limit} to the expected enrichment of the ICM,
since {\it all} the iron not locked in the stars is assumed to be expelled from
the galaxies. Whether this actually occurs, depends on the
efficiency of extraction/ejection mechanisms (feedback, ram pressure, etc.).

The same exercise, described above for iron, can be repeated for any other
element; in particular for the $\alpha$--elements that directly trace SN~II
enrichment and star formation. Among these, silicon is the best measured one 
in the ICM (Fig.~\ref{fig:IMLR_SiMLR_SiFe}b).

From Fig.~\ref{fig:IMLR_SiMLR_SiFe}, our simple argument shows
that with the Salpeter IMF one expects:
\begin{enumerate}
\item
equipartition of iron between stars and ICM 
(Fig.~\ref{fig:IMLR_SiMLR_SiFe}a; Renzini et~al.\ 1993);
\item
an IMLR in the ICM compatible with observations, provided the bulk of the
stellar population in clusters is old (say, older than 10~Gyr;
Fig.~\ref{fig:IMLR_SiMLR_SiFe}a);
\item
an uneven distribution (i.e.\ no equipartition) of $\alpha$--elements, that are
mostly contained in the stars (Fig.~\ref{fig:IMLR_SiMLR_SiFe}b);
\item
a ``chemical asymmetry'' between stars and ICM 
(Fig.~\ref{fig:IMLR_SiMLR_SiFe}c; Matteucci \& Vettolani 1988; 
Renzini et~al.\ 1993; Pipino et~al.\ 2002): the stars have
supersolar [$\alpha$/Fe] abundance ratios --- as we assumed {\it a priori}
after observational results --- while the ICM has undersolar
[$\alpha$/Fe];
the latter result is at odds with observations (Finoguenov et~al.\ 2000, 2001; 
Baumgartner et~al.\ 2004);
\item
a SiMLR lower than observed (Fig.~\ref{fig:IMLR_SiMLR_SiFe}b).
\end{enumerate}
In a hydro--dynamical simulation adopting the Salpeter IMF, these basic
predictions can be used to check and interpret 
the results. For instance, if more iron is contained in the stars 
than in the ICM, it means that the simulated stellar iron abundances are
in excess of the observed ones and a more efficient ejection of iron into the
ICM needs to be implemented. If the IMLR in the simulated ICM is lower than
observed, either not enough iron has been ejected into the ICM,
or the stellar populations are too young (high luminosity); while
a low SiMLR and an undersolar [Si/Fe] ratio in the ICM are
expected with the Salpeter IMF and are not a fault of the simulations.

\begin{figure}
\centering
\resizebox{0.5 \textwidth}{!}{{\includegraphics{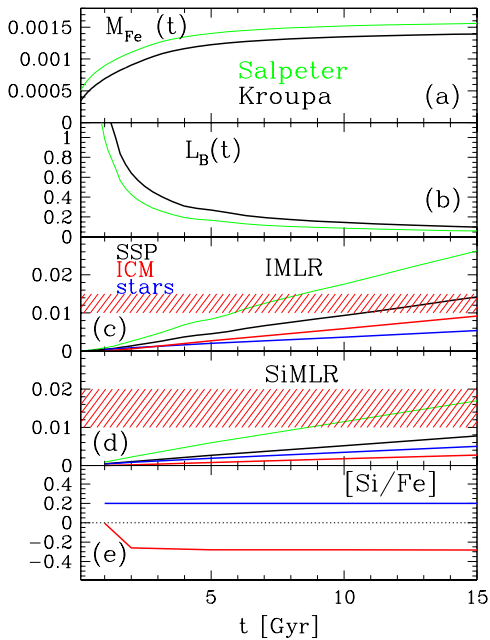}}}
\caption{{\bf (a)} Iron production from a Kroupa SSP, compared to a Salpeter
SSP.
{\bf (b)} Luminosity evolution of a Kroupa vs.\ Salpeter SSP.
{\bf (c)} IMLR$_{SSP}$ for Kroupa and Salpeter;
the Kroupa IMLR$_{SSP}$ is split into its stellar and ICM components,
and the latter is compared to observations (red shaded area).
{\bf (d)} Same as (c), for the SiMLR.
{\bf (d)} [Si/Fe] ratio in the stars and in the ICM, for the Kroupa IMF.}
\label{fig:Kroupa}
\end{figure}

The results listed above refer to the Salpeter IMF, 
but the procedure can be applied to any other IMF; the expected metal
production and partition between stars and ICM will be different.
Fig.~\ref{fig:Kroupa} shows the case for a typical 
Solar Neighbourhood IMF, the Kroupa (1998) IMF. With a steep Scalo slope
at the high--mass end, the Kroupa IMF has a lower iron production than
Salpeter (Fig.~\ref{fig:Kroupa}a). Being also ``bottom--light'' with respect 
to Salpeter (namely, with less mass locked in low--mass stars) it has a higher
luminosity as well (Fig.~\ref{fig:Kroupa}b). All in all, its characteristic 
IMLR$_{SSP}$ is a factor of
2 lower than that of Salpeter, and it cannot possibly match the observed iron 
enrichment in the ICM (Fig.~\ref{fig:Kroupa}c; see PMCS for details). 
Even worse is the outcome for 
the $\alpha$--elements (Fig.~\ref{fig:Kroupa}d). Henceforth, one can
predict beforehand that adopting a ``standard'' Solar Neighbourhood IMF 
in cluster simulations can never lead to a satisfactory ICM enrichment.
Notice also that with the Kroupa IMF there is no
equipartition of iron between stars and ICM (Fig.~\ref{fig:Kroupa}c); 
iron equipartition holds specifically for the Salpeter IMF.

\begin{figure}
\centering
\resizebox{0.35 \textwidth}{!}{{\includegraphics{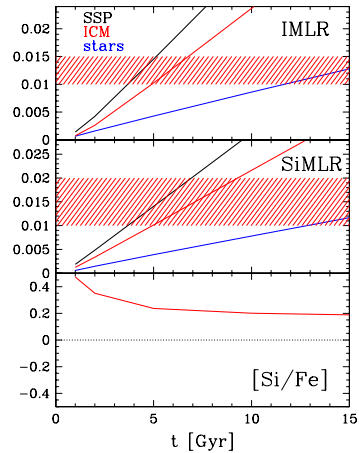}}}
\caption{IMLR, SiMLR and [Si/Fe] ratio in the stars and in the ICM, expected 
from the Arimoto \& Yoshii IMF.}
\label{fig:AY}
\end{figure}

Fig.~\ref{fig:AY} shows the expected production and partition of elements
between stars and ICM, for the top--heavy Arimoto \& Yoshii (1987)
IMF (power--law exponent --1.0 vs.\ --1.35 for Salpeter; same mass limits
[0.1--100]~$M_{\odot}$). For this IMF, more metals are available to enrich the
ICM than locked in the stars, and the IMLR and SiMLR in the ICM can match
the observed levels for stellar populations of 5--10~Gyrs of age. The expected
[$\alpha$/Fe] ratios in the ICM are supersolar, in agreement with observations.
Therefore, a top--heavy IMF like this can potentially lead to satisfactory
results; whether this is achieved in an actual simulation,
depends on the simulated star formation history and ejection
mechanisms of metals from the galaxies.

\section{IMF, M$_*$/L ratio and cold fraction}

In the previous section we showed that a self-consistent account of the mass
and metals locked in the stars, related to the $M_*/L$ corresponding 
to the adopted IMF, is fundamental to predict the chemical enrichment
of the ICM. The $M_*/L$ has also important 
consequences on the cold fraction, defined as the ratio $M_*/M_{ICM}$
between stellar mass and hot ICM gas mass (the cold gas component has a 
negligible mass contribution). The cold fraction is crucial to distinguish, 
for instance,
the mechanisms responsible for the ``entropy floor'' in low temperature 
clusters. The larger the cold fraction, the larger the role played by galaxy 
formation that removes low entropy gas; with a small cold fraction, instead,
strong supernova (pre)heating or other feedback sources are needed to set 
the entropy floor (Bryan 2000; Balogh et~al.\ 2001; Valdarnini 2003;
Tornatore et~al.\ 2003). Consequently, a correct estimate of the cold fraction
is a major constraint for scenarios of cluster evolution.

While $M_{ICM}$ can be inferred directly from the observed
X-ray emission, the stellar mass can be estimated only indirectly from the
observed luminosity, typically in the B--band, via an assumed $M_*/L$.
In literature, very discrepant values can be found for the cold fraction --- 
or its reciprocal:
\[ \frac{M_{ICM}}{M_*} = 2-20 \]
This discrepancy is mostly due to the assumed
$M_*/L$; the determinations of the {\it directly observed} quantity
$\frac{M_{ICM}}{L}$ (listed below as compiled by Moretti et~al.\ 2003) 
are in much better agreement, within 30\% around:
\[ \frac{M_{ICM}}{L_B} = 30 \, h^{-\frac{1}{2}} \frac{M_{\odot}}{L_{\odot}}
                         \simeq 36 \frac{M_{\odot}}{L_{\odot}} 
                         ~~~~~(h=0.7)\]
\smallskip \noindent
\begin{tabular}{l l}
\hline
\multicolumn{2}{c}{$\frac{M_{ICM}}{L_B} \left[ h^{-\frac{1}{2}} 
\frac{M_{\odot}}{L_{\odot}} \right]$} \\
\hline
 30       & David et al.\ (1990), Hydra A \\
 30 ($\pm$ 13) & Arnaud et al.\ (1992) \\
 30 & White et al.\ (1993), Coma \\
 19 & Cirimele et al.\ (1997) \\
 31--44 & Roussel et al. (2000) \\
 $>$21 & Finoguenov et al. (2003) \\
\hline
\end{tabular}

\begin{figure}
\centering
\resizebox{\hsize}{!}{\rotatebox[]{-90}{\includegraphics{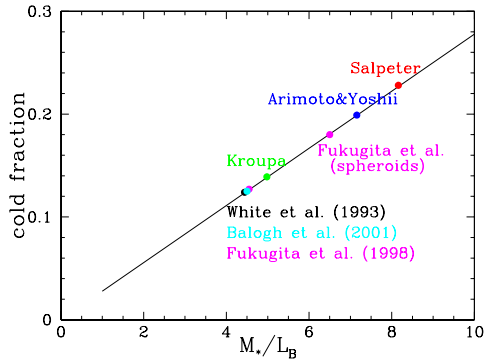}}}
\caption{Cold fraction $M_*/M_{ICM}$ in clusters of galaxies as a function
of the assumed $M_*/L_B$ ratio, for a typical observed value of 
$M_{ICM}/L_B = 36 \, M_{\odot}/L_{\odot}$.}
\label{fig:coldfrac_ML}
\end{figure}

\medskip \noindent
With this typical value of $\frac{M_{ICM}}{L_B}$, the cold 
fraction depends on the assumed $M_*/L_B$ as shown in 
Fig.~\ref{fig:coldfrac_ML}. Displayed
are the values of $M_*/L_B$ assumed by White et al.\ (1993) and Balogh 
et al.\ (2001), by Fukugita et al.\ (1998; averaged over the typical 
morphological mixture of cluster galaxies out to the Abell radius, or for 
spheroids only); and the $M_*/L_B$ values corresponding to the Salpeter, 
Kroupa and Arimoto \& Yoshii IMF, assuming an age of 10~Gyr for the stellar 
populations. The estimated cold fraction ranges between 10 and 20\%;
while such factor of~2 is of little relevance for the ``baryon budget'' 
addressed by White et~al.\ and Fukugita et~al., since the baryonic mass 
in clusters is anyways largely dominated by the hot ICM, it has a drastic 
impact on chemical enrichment, cold fraction and cluster modelling.
Fig.~\ref{fig:coldfrac_ML} shows that, if one assumes a Salpeter 
(or Arimoto \& Yoshii) IMF,
the observed $\frac{M_{ICM}}{L_B}$ corresponds to a cold fraction around 20\%
for suitable ages of the stars in cluster galaxies ($\sim$10~Gyr, suited
to reproduce the IMLR: Fig.~\ref{fig:IMLR_SiMLR_SiFe}a).
The widely quoted cold fraction of 10\% can be reached only by simulations
that assume an IMF corresponding to $M_*/L_B$ values as low as assumed by 
e.g.\ White et~al.\ (1993).

The correct constraint for a simulation is the cold fraction
{\it corresponding to the assumed IMF}, based on the {\it observed}
$\frac{M_{ICM}}{L_B}$. Equivalently, for the simulated cluster one can 
compute self--consistently the luminosity relevant to the assumed IMF 
and compare directly to the observed $\frac{M_{ICM}}{L_B}$.

Alternatively, if simulations aim at reaching a cold fraction of 10\%,
one should select beforehand an IMF (quite ``bottom light'') able
to reach as low values of $M_*/L_B$ as assumed by White et al., for the
old ages typical of elliptical galaxies.

\section{Conclusions}
We remark the following crucial points for modelling the cosmological evolution
and chemical enrichment of clusters of galaxies self--consistently.
\begin{itemize}
\item
The amount of mass {\it and} metals locked in the stellar component is not 
necessarily negligible, depending on the assumed IMF and corresponding $M_*/L$.
\item
The observed metal content in stars can be used as a constraint for the 
efficiency of feed--back and of metal dispersion in the simulations: the wind 
efficiency cannot be indefinitively enhanced, because the amount of metals 
in the stars should also be accounted for, and a sizable fraction of the 
metals produced may thus not be available to enrich the ICM. (Though the 
problem of present--day simulations is rather the opposite, namely to lock 
too much metals in the stars; Tornatore et~al.\ 2004; Romeo et~al.\ 2004).
\item
Once an IMF is chosen for the simulations (preferably a bottom--light IMF 
with high-mass slope shallower than Scalo, see PMCS), we suggest
to compute the corresponding partition of metals between stars and ICM 
with the simple procedure outlined in this paper. Such expected partition
can then be used to test the numerical results.
\item
The crucial constraint of the cold fraction should be consistent with
the $M_*/L$ relevant to the assumed IMF. As the
observational quantity is luminosity, we suggest to compute the luminosity
of the simulated cluster consistently with the adopted IMF, and compare it
directly to the
{\it observed} $M_{ICM}/L$. Red or IR luminosities are preferred,
since they probe the stellar mass with a lower sensitivity to the age of the 
stellar populations and to recent star formation activity; $M_{ICM}/L_K$ 
estimates are becoming available (Lin et~al.\ 2003).
\end{itemize}
\bibliographystyle{aa}

\end{document}